\documentclass[sigconf,noacm]{acmart}

\usepackage{amsmath}
\usepackage{booktabs}
\usepackage{graphicx}
\usepackage{xspace}
\usepackage{tikz}
\usepackage{tabularx}
\usetikzlibrary{arrows.meta,positioning}

\renewcommand\footnotetextcopyrightpermission[1]{}

\newcommand{\method}{AnchorQE\xspace}
\newcommand{\trec}{TREC-DL\xspace}
\newcommand{\lotte}{LoTTE\xspace}

\newcommand{\ndcg}{nDCG@10\xspace}
\newcommand{\success}{Success@5\xspace}
\newcommand{\Eq}{E_q}
\newcommand{\Ed}{E_d}
\newcommand{\norm}[1]{\left\lVert#1\right\rVert_2}
\newcommand{\best}[1]{\textbf{#1}}
\begin{document}

\title{Query Expansion Is More Than Generation: Improving Dense Retrieval through Better Integration}

\author{Siyuan Sun}
\email{wilsonsun@arizona.edu}
\affiliation{%
  \institution{University of Arizona}
  \city{Tucson}
  \state{AZ}
  \country{USA}
}

\author{Mihai Surdeanu}
\email{msurdeanu@arizona.edu}
\affiliation{%
  \institution{University of Arizona}
  \city{Tucson}
  \state{AZ}
  \country{USA}
}

\begin{abstract}
Large language models (LLMs) can generate query expansions without task-specific training, yet the same expansions often make a frozen dense retriever worse. We identify an underexplored factor: prior work has often focused on what text is generated, while how generated text is incorporated into dense retrievers has received less systematic attention. By holding generated expansions fixed, we show that performance degradation can often be attributed to the integration method itself. We introduce AnchorQE, a training-free method that separately encodes the original query and its expansion before interpolating them. The interpolation factor is estimated using an unsupervised online strategy that operates over a small part of the unlabeled test stream. Intuitively, our strategy assigns high expansion trust only when expansions are both retrieval-strong and consistent with the original query's retrieved evidence. We show that AnchorQE improves retrieval effectiveness by up to 12.89\% when compared to widely-used expansion-only or text-level concatenation baselines across TREC-DL, LoTTE, and BEIR. Further, we show that our online strategy to estimate the interpolation factor outperforms a fixed weight tuned on a development partition by up to 3.81\%.
\end{abstract}


\maketitle

\pagestyle{plain}

\section{Introduction}

\begin{figure*}[!t]
\centering
\begin{tikzpicture}[
  font=\footnotesize,
  >=Latex,
  source/.style={draw,rounded corners,fill=black!4,minimum height=6mm,
    inner xsep=6pt,align=center},
  route/.style={draw,rounded corners,minimum width=.225\textwidth,
    minimum height=21mm,inner sep=5pt,align=center},
  note/.style={font=\scriptsize,align=center,text=black!70},
  every path/.style={line width=.45pt}
]
  \node[source] (fixed) {\textbf{Fixed upstream and backend:}\quad
    query $q$\quad$+$\quad saved LLM expansion $z=G(q)$\quad$+$\quad
    frozen $E_q,E_d$, and document index};

  \node[route,below=8mm of fixed,xshift=-.36\textwidth,
        fill=red!3,draw=red!45!black] (exp) {
    \textbf{Expansion only}\\[-1pt]
    $z\rightarrow E_q\rightarrow\mathbf z\rightarrow\mathrm{Index}$};
  \node[route,right=4mm of exp,fill=red!3,draw=red!45!black] (text) {
    \textbf{Text re-encoding}\\[-1pt]
    $[q;\mathrm{sep};z]\rightarrow E_q\rightarrow\mathbf v$\\[-1pt]
    $\mathbf v\rightarrow\mathrm{Index}$};
  \node[route,right=4mm of text,fill=orange!5,draw=orange!65!black] (fusion) {
    \textbf{Separate score fusion}\\[-1pt]
    $q\rightarrow E_q\rightarrow\mathrm{Index}$\\[-2pt]
    $z\rightarrow E_q\rightarrow\mathrm{Index}$\\[-1pt]
    $(1-\alpha)s_q+\alpha s_z$};
  \node[route,right=4mm of fusion,fill=blue!6,draw=blue!60!black] (anchorqe) {
    \textbf{AnchorQE / SC-AnchorQE}\\[-1pt]
    $q,z\xrightarrow{\;E_q\;}\mathbf q,\mathbf z$\\[-1pt]
    $\widehat{\mathbf q}_{\alpha}=
      \mathrm{norm}((1-\alpha)\mathbf q+\alpha\mathbf z)$\\[-1pt]
    SC: $\alpha$ from an unlabeled stream prefix};

  \draw[->] (fixed.south) -- ++(0,-3mm) -| (exp.north);
  \draw[->] (fixed.south) -- ++(0,-3mm) -| (text.north);
  \draw[->] (fixed.south) -- ++(0,-3mm) -| (fusion.north);
  \draw[->] (fixed.south) -- ++(0,-3mm) -| (anchorqe.north);

  \node[note,below=1.5mm of exp] {query dropped; implicit trust\\
    one logical request};
  \node[note,below=1.5mm of text] {query retained but entangled\\
    implicit trust; one request};
  \node[note,below=1.5mm of fusion] {query retained in a separate route\\
    explicit trust; two requests + merge};
  \node[note,below=1.5mm of anchorqe] {query anchored in one vector\\
    explicit trust; one logical request};
\end{tikzpicture}
\caption{Experimental overview of generation and integration in dense QE. Every route receives the same query and saved expansion and uses the same frozen retriever; only the integration method changes. Published pipelines primarily occupy the first two routes. Separate score fusion exposes trust but requires multiple retrieval streams, whereas AnchorQE compiles the same homogeneous linear-score objective into one query-anchored vector and one logical index request. SC-AnchorQE estimates the interpolation factor from a short unlabeled stream prefix and freezes it for future queries.}
\Description{A fixed query, saved LLM expansion, query encoder, document encoder, and index feed four alternative integration routes: expansion only, joint text re-encoding, separate two-stream score fusion, and AnchorQE or SC-AnchorQE. Annotations compare whether the query is anchored, whether expansion trust is explicit, and whether one or two logical index requests are required.}
\label{fig:overview}
\end{figure*}
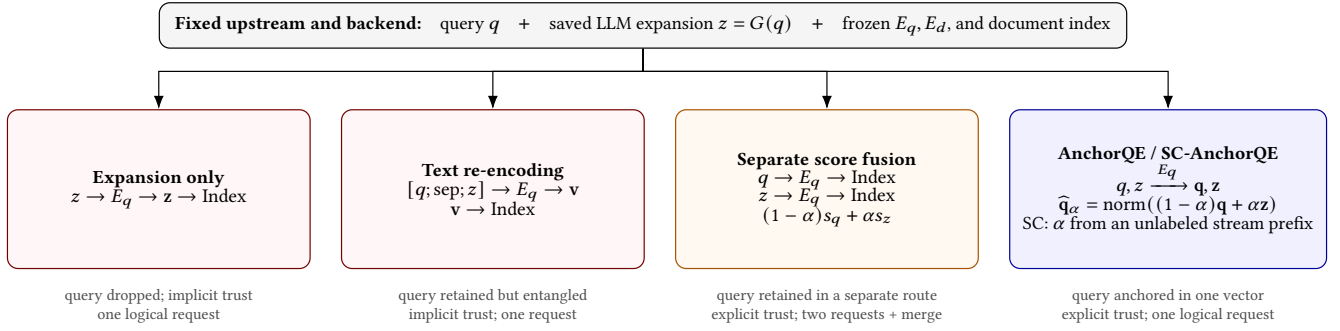

Query expansion (QE) is a natural use of large language models: an LLM can turn a short query into a hypothetical document, a pseudo-document, or related terms without task-specific training \cite{gao2023hyde,wang2023query2doc,jagerman2023query,zhang-etal-2024-exploring-best}. Yet these methods are not reliably beneficial with a frozen dense retriever. 
Our preliminary analysis showed that conventional integration methods fall below the dense-retrieval baseline (DR baseline) without query expansion on multiple datasets (see Figure~\ref{fig:motivation}).
The first three blocks in Figure~\ref{fig:overview} summarize the controlled integration routes averaged here.
Because only the integration method changes, the degradation cannot be attributed to generation quality alone.

\begin{figure}[!t]
\centering
\begin{tikzpicture}[x=.62cm,y=.56cm,font=\scriptsize]
  \draw[->,black!55] (-5.5,0) -- (2.6,0)
    node[right] {$\Delta$ vs.\ DR baseline (points)};
  \foreach \x in {-5,-4,-3,-2,-1,0,1,2} {
    \draw[black!25] (\x,-.12) -- (\x,.12);
    \node[below=2pt,black!65] at (\x,-.12) {\x};
  }
  \draw[black!35,dashed] (0,.25) -- (0,5.45);
  \node[left] at (-5.25,4.8) {T19};
  \node[left] at (-5.25,3.8) {T20};
  \node[left] at (-5.25,2.8) {Search};
  \node[left] at (-5.25,1.8) {Forum};
  \node[left] at (-5.25,.8) {BEIR-14};
  \fill[blue!65] (0,4.5) rectangle (1.52,5.1);
  \fill[red!68] (-3.28,3.5) rectangle (0,4.1);
  \fill[red!68] (-2.92,2.5) rectangle (0,3.1);
  \fill[red!68] (-4.13,1.5) rectangle (0,2.1);
  \fill[red!68] (-4.53,.5) rectangle (0,1.1);
  \node[right] at (1.52,4.8) {$+1.52$};
  \node[right,text=white] at (-3.28,3.8) {$-3.28$};
  \node[right,text=white] at (-2.92,2.8) {$-2.92$};
  \node[right,text=white] at (-4.13,1.8) {$-4.13$};
  \node[right,text=white] at (-4.53,.8) {$-4.53$};
\end{tikzpicture}
\caption{Empirical evidence that zero-shot QE is not robust under conventional integration. 
Bars show the changes in retrieval performance caused by conventional QE using its original integration method, averaged over four QE strategies.
Each row indicates a different dataset (see 4.2 for details).
T19/T20 and BEIR-14 use \ndcg; LoTTE uses \success. 
}
\Description{A horizontal bar chart shows a 1.52-point gain on TREC 2019 and losses of 2.92 to 4.53 points on TREC 2020, LoTTE Search, LoTTE Forum, and BEIR-14. Negative values are printed inside the bars, and the positive value is printed beside its bar.}
\label{fig:motivation}
\end{figure}
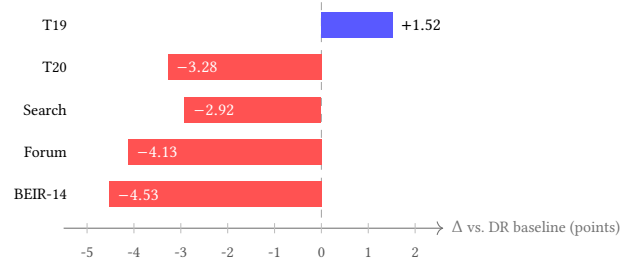

\begin{table*}[t]
\centering
\caption{Representative LLM-QE integration methods. ``Explicit factor'' means that an interpolation or fusion factor controls expansion contribution. Requests count conventional index-side retrievals after encoding.}
\label{tab:prior}
\small
\setlength{\tabcolsep}{3.5pt}
\begin{tabular}{p{.13\textwidth}p{.19\textwidth}p{.29\textwidth}
                p{.11\textwidth}p{.11\textwidth}p{.08\textwidth}}
\toprule
Method & Generated object & Integration method & Query anchor &
Explicit factor & Requests \\
\midrule
HyDE \cite{gao2023hyde} & Hypothetical document &
Expansion-only or vector mean & None & Coupled to averaging & 1 \\
Query2Doc \cite{wang2023query2doc} & Pseudo-document &
Query--document text re-encoding & In text & No & 1 \\
Prompted QE \cite{jagerman2023query} & Keywords / CoT terms &
Repeated-query text re-encoding & In text & No & 1 \\
MuGI \cite{zhang-etal-2024-exploring-best} & Multiple pseudo-references &
Contextual feature pooling & In re-encoding & No & 1 \\
Exp4Fuse \cite{liu2025exp4fuse} & Hypothetical document &
Modified reciprocal-rank fusion & Separate route & Fusion-specific & 2 \\
MMLF \cite{kuo2025mmlf} & Subqueries + passages &
Reciprocal-rank fusion & Separate route & No & $N+1$ \\
\method (ours) & Any embeddable expansion &
Normalized vector interpolation & In one vector & Yes & 1 \\
\bottomrule
\end{tabular}
\end{table*}

The overlooked variable is the \emph{integration method}. Expansion-only retrieval discards the original query; text concatenation retains its tokens but entangles query and expansion through tokenization, attention, pooling, and truncation. Neither method explicitly controls how far generated evidence can move the dense query representation. We therefore propose \method, a framework to separately encode the original query and expansion, then interpolate their normalized vectors. This \emph{text-level expansion and vector-level integration} paradigm is designed for zero-shot QE with a frozen dense retriever: it accepts any expansion that the query encoder can embed, requires neither retraining nor an index rebuild, and produces one query vector for a standard dense-index request. The original query remains an explicit anchor.

The remaining research question is how much to trust an expansion on a new stream without relevance labels. We introduce Stream-Calibrated AnchorQE (SC-AnchorQE), which estimates the interpolation factor from the first few unlabeled queries (eight in our experiments) and then freezes it for future queries. Expansion trust is high only when the expansion retrieves strongly on its own and agrees with evidence retrieved by the original query. No training, relevance judgments, or future-query effectiveness are used.

Our contributions are:
\begin{itemize}
\item We isolate integration as an experimental variable with a fixed-expansion protocol covering four LLM generation strategies. The resulting comparison shows that conclusions about QE can change solely with the integration method.
\item We introduce \method, a training-free, index-preserving integration framework that keeps the query explicit, makes expansion trust controllable, and retains one-vector serving. We prove equivalence to homogeneous weighted score fusion and a worst-case angular influence bound.
\item We propose a label-free online strategy for estimating the interpolation factor and validate it across TREC-DL, LoTTE, and BEIR \cite{bajaj2016msmarco,craswell2020overview,craswell2021overview,santhanam2022lotte,thakur2021beir}, against the DR baseline, text re-encoding, and reproduced QuDAR baselines, and across nine generator--retriever configurations. 
\end{itemize}
\typeout{WSDM-CONTRIBUTIONS-END-PAGE=\thepage}

\section{Previous Work and Preliminaries}
\label{sec:background}

\subsection{Dense Retrieval}

Let $\mathcal D$ be a document collection. A dual-encoder dense retriever maps a query and document to vectors:
\[
  \mathbf q=\operatorname{normalize}(\Eq(q)),\qquad
  \mathbf d=\operatorname{normalize}(\Ed(d)),
\]
and ranks by $s(q,d)=\mathbf q^\top\mathbf d$ \cite{karpukhin2020dpr,izacard2022contriever}. The functions $\Eq$ and $\Ed$ may share parameters, but modern embedding models still distinguish roles through task instructions or prefixes. QE must therefore specify not only the generated text but also how it enters the query-side representation.

\subsection{Query Expansion}

Automatic QE traditionally augments a query with related terms or feedback evidence \cite{carpineto2012survey}. LLMs broaden the generated object. HyDE searches with hypothetical-document embeddings \cite{gao2023hyde}; Query2Doc appends a pseudo-document to the query \cite{wang2023query2doc}; prompted QE generates keywords, subqueries, or reasoning-derived terms \cite{jagerman2023query,zhang-etal-2024-exploring-best}. We write $z=G(q)$ and $\mathbf v=I(q,z)$, where generation strategy $G$ includes the LLM, prompt, and decoding, while integration method $I$ maps the query and saved expansion to the representation used for retrieval.

Table~\ref{tab:prior} separates these two decisions for representative LLM-QE systems and records whether integration retains an explicit query anchor, exposes a trust factor, and requires additional index requests.

Existing work primarily varies what to generate and how many samples to draw. Because generation and integration usually change together, their effects are confounded. Our fixed-expansion protocol saves each output once, reuses it across methods, and changes only $I$.

\subsection{Score Fusion}

An alternative is to retrieve separately with the query and expansion, then combine scores or ranks. Weighted CombSUM \cite{shaw1994combination} assigns
\begin{equation}
 s_{\mathrm{sum}}(d)=(1-\alpha)s_q(d)
   +\alpha\sum_{i=1}^{N}\pi_i s_{z_i}(d),
 \label{eq:combsum}
\end{equation}
where $\pi_i\geq0$ and $\sum_i\pi_i=1$. Reciprocal rank fusion (RRF) instead combines truncated ranks, typically as $s_{\mathrm{rrf}}(d)=\sum_i(k+r_i(d))^{-1}$ \cite{cormack2009rrf}. Score fusion exposes weights, but a conventional implementation submits $N+1$ index requests and merges candidate lists. \method recovers the linear objective of Eq.~\ref{eq:combsum} in one vector when all streams share the same document space.

\subsection{Prior Work on QE Integration}

MuGI \cite{zhang-etal-2024-exploring-best} is one of the few LLM-QE studies that makes integration a central design concern. It samples multiple pseudo-references, balances them with the query for sparse retrieval, and studies contextualized pooling and pseudo-relevance-feedback calibration for dense retrieval. Its research question is how multi-text generation can be made broadly effective across retriever families. Generation multiplicity, retriever choice, and integration consequently form one framework, rather than a controlled comparison in which the expansion text is fixed and only its entry into a dense retriever changes.

QuDAR \cite{kim-etal-2026-qudar} studies a different integration layer: query-wise hybrid retrieval across retriever type (sparse versus dense) and query format (original versus expanded). It obtains four ranked lists and combines them with RRF, equal score fusion, margin-derived confidence, or LLM-based relevance weights. The central problem is adaptive fusion of heterogeneous retrieval outputs. AnchorQE instead addresses integration inside one frozen dense space, where original-query and expansion evidence must become one query vector with an explicit trust factor. This difference is architectural---post-retrieval multi-stream fusion versus pre-retrieval single-vector integration---rather than simply a different weighting formula. Section~\ref{sec:qudar} reproduces QuDAR's released fusion rules on the same expansion texts and collections for a direct comparison.

\section{AnchorQE}
\label{sec:method}

\subsection{Definition}

For a query $q$ and expansions $z_1,\ldots,z_N$, \method separately encodes and normalizes the query and each expansion:
\[
 \mathbf q=\operatorname{normalize}(\Eq(q)),\qquad
 \mathbf z_i=\operatorname{normalize}(\Eq(z_i)).
\]
Let $\pi_i\geq0$ and $\sum_i\pi_i=1$. \method first forms the expansion mixture $\mathbf m=\sum_{i=1}^{N}\pi_i\mathbf z_i$, then submits
\begin{equation}
 \widehat{\mathbf q}_{\alpha}=
 \operatorname{normalize}\!\left(
   (1-\alpha)\mathbf q+\alpha\mathbf m\right),
 \qquad 0\leq\alpha\leq1.
 \label{eq:anchorqe}
\end{equation}
unchanged.
For $N=1$, Eq.~\ref{eq:anchorqe} interpolates the query with one expansion. For $N>1$, $\alpha$ controls the total contribution of the mixture. With no expansion or $\alpha=0$, \method returns $\mathbf q$; $\Ed$ and the index remain unchanged.

This construction has three direct consequences. The deployed vector retains the original query explicitly; $\alpha$ controls expansion influence independently of expansion length; and disagreement among multiple expansion vectors reduces $\norm{\mathbf m}$, weakening their joint correction. These properties are explicit in vector interpolation but are not guaranteed by encoder-dependent text re-encoding.

\subsection{Unsupervised Online Factor Estimation}

SC-AnchorQE estimates one interpolation factor from an unlabeled prefix of $B$ queries and freezes it for the remaining stream. We use one expansion per query. For prefix query $i$, let $d^q_{i,1}$ and $d^z_{i,1}$ be the top-1 documents retrieved by $\mathbf q_i$ and $\mathbf z_i$, and let $D^q_{i,10}$ be the original query's top-10. With $[x]_+=\max(x,0)$, define
\begin{align}
s_{q,1}&=\tfrac1B\sum_i[\mathbf q_i^\top\mathbf d^q_{i,1}]_+,&
s_{z,1}&=\tfrac1B\sum_i[\mathbf z_i^\top\mathbf d^z_{i,1}]_+,\nonumber\\
s_{q,\mathrm{sup}}&=\tfrac1B\sum_i
 [\tfrac1{10}\!\sum_{\mathbf d\in D^q_{i,10}}\!\mathbf q_i^\top\mathbf d]_+,&
s_{z,\mathrm{sup}}&=\tfrac1B\sum_i
 [\tfrac1{10}\!\sum_{\mathbf d\in D^q_{i,10}}\!\mathbf z_i^\top\mathbf d]_+.
\label{eq:signals}
\end{align}
The first ratio measures the expansion's relative retrieval strength; the second measures its agreement with evidence retrieved by the original query:
\begin{equation}
r_{\mathrm{top1}}=\frac{s_{z,1}}{s_{q,1}+s_{z,1}},\quad
r_{\mathrm{sup}}=\frac{s_{z,\mathrm{sup}}}
 {s_{q,\mathrm{sup}}+s_{z,\mathrm{sup}}}.
\label{eq:ratios}
\end{equation}
SC-AnchorQE uses their conjunctive product
\begin{equation}
\boxed{\alpha_{\mathrm{stream}}=r_{\mathrm{top1}}r_{\mathrm{sup}}.}
\label{eq:stream-alpha}
\end{equation}
The product assigns substantial expansion trust only when both conditions hold. This parameter-free product directly implements that conjunctive intuition and was designed without reference to benchmark performance. It has no fitted parameters, thresholds, relevance labels, or access to future-query effectiveness. Each generation strategy is calibrated independently. During prefix calibration, each query needs two probe retrievals---one with $\mathbf q_i$ and one with $\mathbf z_i$---to obtain the top documents in Eq.~\ref{eq:signals}. These probes are calibration-only; after $\alpha$ is frozen, every future query uses the single AnchorQE request in Eq.~\ref{eq:anchorqe}.

\subsection{Equivalence to Linear Score Fusion}

\paragraph{Proposition 1.} Assume query and expansion streams score the same document vectors by raw dot product and $(1-\alpha)\mathbf q+\alpha\mathbf m\neq\mathbf0$. Then exact full-corpus retrieval with $\widehat{\mathbf q}_{\alpha}$ has the same document ordering as weighted CombSUM with weights $(1-\alpha,\alpha\pi_1,\ldots,\alpha\pi_N)$.

\paragraph{Proof.} For every document vector $\mathbf d$,
\begin{align}
 \widehat{\mathbf q}_{\alpha}^{\top}\mathbf d
 =\frac{(1-\alpha)\mathbf q^\top\mathbf d+
 \alpha\sum_i\pi_i\mathbf z_i^\top\mathbf d}
 {\norm{(1-\alpha)\mathbf q+\alpha\mathbf m}}.
 \label{eq:equivalence}
\end{align}
The nonzero assumption makes the denominator positive; because it is also independent of $d$, it cannot change the ranking. The numerator is exactly Eq.~\ref{eq:combsum}. \hfill$\square$

Note that this is a ranking equivalence, not equality of calibrated scores. It does not apply to mixed sparse--dense streams, stream-specific score normalization, non-linear fusion, or independently truncated ANN candidate lists. 

However, \method has one important advantage: it issues only one retrieval request (because all representations are compiled into a single vector), whereas CombSUM uses $N+1$ retrieval requests.

\begin{table*}[t]
\centering
\caption{Fixed-expansion integration diagnostic. T19/T20/BEIR report \ndcg; LoTTE reports \success. The DR baseline is a shared reference. Within each QE-strategy block, bold marks the best integration method for each benchmark. This table shows that conventional integration methods fall below DR baseline, while our proposed framework improves retrieval performance by a large margin.}
\label{tab:main}
\scriptsize
\setlength{\tabcolsep}{3.8pt}
\begin{tabular}{llccccc}
\toprule
QE strategy & Integration method & T19 & T20 & Search & Forum & BEIR-14 \\
\midrule
None & DR baseline & .6765 & .7056 & .7780 & .7470 & .5387 \\
\midrule
HyDE & Published & .6693 & .6931 & .7490 & .6916 & .4892 \\
HyDE & Tuned text & \best{.7169} & .6879 & .7701 & .7175 & .5123 \\
HyDE & AnchorQE $.15$ & .7154 & \best{.7303} & \best{.7913} & \best{.7573} & \best{.5486} \\
\addlinespace
Query2Doc & Published & .7267 & .7128 & .7737 & .7166 & .5107 \\
Query2Doc & Tuned text & .7191 & .6878 & .7632 & .7073 & .5113 \\
Query2Doc & AnchorQE $.15$ & \best{.7300} & \best{.7299} & \best{.7924} & \best{.7569} & \best{.5472} \\
\addlinespace
Q2E & Published & .6941 & .6478 & .7449 & .7128 & .4920 \\
Q2E & Tuned text & .7098 & .6571 & .7603 & .7256 & .5138 \\
Q2E & AnchorQE $.15$ & \best{.7236} & \best{.7117} & \best{.7853} & \best{.7519} & \best{.5437} \\
\addlinespace
CoT terms & Published & .6766 & .6374 & .7275 & .7018 & .4816 \\
CoT terms & Tuned text & .6976 & .6570 & .7597 & .7289 & .5147 \\
CoT terms & AnchorQE $.15$ & \best{.7248} & \best{.7161} & \best{.7863} & \best{.7542} & \best{.5437} \\
\bottomrule
\end{tabular}
\end{table*}

\subsection{Why the Anchor Controls Expansion Influence}

\paragraph{Proposition 2.} Let $\phi_\alpha=\angle(\mathbf q,\widehat{\mathbf q}_{\alpha})$, where $\angle$ indicates the angle between the original query vector and the vector after \method expansion. Then, for interpolation factors between 0 and $1/2$, i.e., $0\leq\alpha<1/2$:
\begin{equation}
 \phi_\alpha\leq
 \arcsin\!\left(\frac{\alpha}{1-\alpha}\right).
 \label{eq:cap}
\end{equation}

\paragraph{Proof.} Because $\norm{\mathbf m}\leq1$, every unnormalized AnchorQE vector lies in a ball of radius $\alpha$ centered at $(1-\alpha)\mathbf q$. For $\alpha<1/2$ the origin lies outside this ball. The largest ray angle from $\mathbf q$ intersecting the ball is the tangent angle, whose sine is the radius divided by the center distance. \hfill$\square$

For example, at $\alpha=.10$, no expansion can alter the retrieval direction by more than $6.38^\circ$, independent of its content. The bound controls influence rather than relevance. A wrong expansion can still cross a small ranking margin, while an informative expansion may deserve more trust.

The same limitation can be stated directly for a document pair. Let $m_q=\mathbf q^\top(\mathbf d_i-\mathbf d_j)$ and $m_z=\mathbf z^\top(\mathbf d_i-\mathbf d_j)$. For $N=1$, the ordering has the sign of $(1-\alpha)m_q+\alpha m_z$. If $m_q>0$ and $m_z<0$, reversal requires
\[
 |m_z|>\frac{1-\alpha}{\alpha}m_q.
\]
For example, at $\alpha=.10$, the opposing expansion margin must exceed the query margin by a factor of nine.

Informally, this proof demonstrates that a conflicting expansion must express a much stronger preference than the original query before it can reverse the order of two documents. The anchor therefore limits how easily an expansion can overturn the query, without assuming that the original order is always correct.

\section{Experimental Setup}
\label{sec:setup}

\begin{table*}[t]
\centering
\caption{
Primary result of SC-AnchorQE. Each cell reports the SC-AnchorQE score followed by its absolute delta versus the published-integration baseline. Our method outperforms conventional QE integration methods in all cases.
}
\label{tab:prefix}
\small
\setlength{\tabcolsep}{5pt}
\begin{tabular}{lccccc}
\toprule
QE strategy & T19 & T20 & Search & Forum & BEIR-14 \\
\midrule
HyDE & .7346 $(+.0518)$ & .7292 $(+.0215)$ & .7981 $(+.0187)$ & .7612 $(+.0408)$ & .5523 $(+.0358)$ \\
Query2Doc & .7560 $(+.0354)$ & .7340 $(+.0125)$ & .7975 $(+.0235)$ & .7592 $(+.0427)$ & .5498 $(+.0385)$ \\
Q2E & .7319 $(+.0259)$ & .6973 $(+.0398)$ & .7863 $(+.0412)$ & .7545 $(+.0416)$ & .5468 $(+.0524)$ \\
CoT terms & .7354 $(+.0552)$ & .7074 $(+.0543)$ & .7897 $(+.0624)$ & .7538 $(+.0521)$ & .5465 $(+.0618)$ \\
\bottomrule
\end{tabular}
\end{table*}

\subsection{Model Selection}

Our primary generator is Qwen3-8B with greedy decoding, thinking disabled, seed 42, and at most 128 new tokens \cite{yang2025qwen3}. Four generation strategies span document-like and term-like expansions: HyDE, Query2Doc, Q2E keywords, and chain-of-thought-derived terms. Each expansion is saved once and reused across every integration method under comparison. Generator transfer uses Qwen3-1.7B and Llama-3.1-8B-Instruct \cite{grattafiori2024llama3}.

The primary retriever is BGE-large-en-v1.5 \cite{xiao2023cpack}. We select GTE large-en-v1.5 \cite{li2023gte} as a second conventional dual encoder from a different training family, and Qwen3-Embedding-0.6B \cite{zhang2025qwen3embedding} as a decoder-derived embedding model with a different architecture and instruction format. All model-specific query prefixes are preserved; all document embeddings and indexes are frozen. Cross-family generators test whether the result depends on one LLM's lexical style, while cross-architecture retrievers test whether it depends on one encoder geometry. The transfer matrix is the complete Cartesian product of three generators (Qwen3-8B, Qwen3-1.7B, and Llama-3.1-8B) and three retrievers (BGE, GTE, and Qwen3-Embedding), for nine configurations in total.

\subsection{Dataset Selection and Metrics}

We use 18 benchmark streams organized into five reporting groups. \trec 2019 and 2020 provide human judgments for MS MARCO multi-passage queries \cite{bajaj2016msmarco,craswell2020overview,craswell2021overview}. \lotte contributes Search and Forum queries in writing, recreation, science, technology, and lifestyle \cite{santhanam2022lotte}. BEIR-14 contains 14 heterogeneous datasets \cite{thakur2021beir}; CQADupStack is macro-averaged over its 12 domains and contributes one BEIR stream.

This combination is deliberately broad. TREC-DL measures in-domain, deeply-judged passage ranking; LoTTE covers long-tail search and forum questions; BEIR spans biomedical IR, question answering, argument retrieval, duplicate questions, citation prediction, entity retrieval, and fact checking. TREC-DL and BEIR use per-query \ndcg. LoTTE uses per-query \success. We first average queries within each collection, then macro-average the five LoTTE domains or 14 BEIR datasets. We never pool raw queries across collections.

\subsection{Baselines and Evaluation Protocol}

The \emph{published-integration baseline} pairs each generation strategy with the integration method used in its source pipeline. It uses expansion-only for HyDE, query-repeat-five text re-encoding for Q2E and CoT-derived terms, and the following Query2Doc method:
\[
  \operatorname{normalize}\!\left(\Eq(q\,[\mathrm{SEP}]\,z)\right).
\]
This baseline reproduces the integration procedure specified by each source paper while using the shared saved expansions and frozen retriever required by our controlled comparison. For a stronger text baseline, each QE strategy chooses among 72 concatenation and re-encoding recipes on the separate 6,980-query MS MARCO development set (see Appendix C for more details). The same development set selects $\alpha=.15$ for the fixed-factor AnchorQE comparison. All final methods receive identical saved expansions.

We use $B$ to denote the number of unlabeled calibration queries at the start of a stream and set $B=8$.
For each QE strategy, we take the first eight unlabeled queries of a stream and use only those queries to estimate Eq.~\ref{eq:stream-alpha}. We then freeze the interpolation factor, remove the eight calibration queries from evaluation, and score every method on the same remaining suffix. The stored natural order defines this strict no-lookahead split. TREC years and BEIR datasets are separate streams; each LoTTE group is one stream spanning five domains. The DR baseline, fixed AnchorQE, development-tuned text re-encoding, and SC-AnchorQE are always scored on exactly the same future suffix. The full-query matrix used to compare all baselines is explicitly a supporting integration diagnostic, not the primary online test.

Each QE strategy is an independent experimental condition: expansions, calibration signals, interpolation factors, scores, and inference are never pooled across strategies. For each strategy separately, we run 10,000 paired bootstrap replicates within each physical collection while preserving the LoTTE and BEIR nested macros, and report unique-query wins, losses, and ties. At collection level, we use 100,000 cluster-bootstrap replicates over the 18 logical collections and an exact paired sign randomization test. As a secondary arrival-order robustness check, we repeat prefix-to-future evaluation for 20 random orders. Their positive-cell counts and standard deviations are not used as the main significance evidence.

\section{Main Results}
\label{sec:results}

\begin{table*}[t]
\centering
\caption{
Statistical analysis for SC-AnchorQE against the DR baseline, published-integration baseline, and fixed-factor \method ($\alpha=.15$), under four QE strategies. Results show that SC-AnchorQE significantly outperforms both the DR baseline and published integration, while matching or improving a development-tuned fixed factor without relevance labels.
}
\label{tab:inference}
\scriptsize
\setlength{\tabcolsep}{3.3pt}
\begin{tabular}{llccrrr}
\toprule
QE strategy & Comparator & Positive groups & Query CI & Query W/L/T & Collection $\Delta$ [95\% CI] & Sign $p$ \\
\midrule
HyDE & DR baseline & 5/5 & 5/5 & 7,096/4,791/48,339 & +.0172 [+.0096,+.0262] & .000168 \\
HyDE & Published & 5/5 & 3/5 & 14,132/5,694/40,400 & +.0352 [+.0215,+.0495] & $7.63\!\times\!10^{-5}$ \\
HyDE & Fixed $.15$ & 5/5 & 4/5 & 3,897/3,261/53,068 & +.0045 [+.0018,+.0078] & .00311 \\
\addlinespace
Query2Doc & DR baseline & 5/5 & 5/5 & 6,823/4,703/48,700 & +.0166 [+.0082,+.0278] & .000153 \\
Query2Doc & Published & 5/5 & 3/5 & 14,076/5,510/40,640 & +.0363 [+.0228,+.0506] & $2.29\!\times\!10^{-5}$ \\
Query2Doc & Fixed $.15$ & 5/5 & 4/5 & 3,693/3,159/53,374 & +.0038 [+.0011,+.0074] & .00849 \\
\addlinespace
Q2E & DR baseline & 4/5 & 4/5 & 5,873/5,117/49,236 & +.0100 [+.0032,+.0190] & .00409 \\
Q2E & Published & 5/5 & 4/5 & 15,608/5,191/39,427 & +.0490 [+.0282,+.0728] & $3.05\!\times\!10^{-5}$ \\
Q2E & Fixed $.15$ & 4/5 & 3/5 & 3,184/3,218/53,824 & +.0015 [$-.0006$,+.0039] & .211 \\
\addlinespace
CoT terms & DR baseline & 5/5 & 4/5 & 5,835/5,329/49,062 & +.0106 [+.0034,+.0201] & .00375 \\
CoT terms & Published & 5/5 & 4/5 & 16,197/5,061/38,968 & +.0605 [+.0402,+.0833] & $7.63\!\times\!10^{-6}$ \\
CoT terms & Fixed $.15$ & 3/5 & 1/5 & 3,110/3,406/53,710 & +.0014 [$-.0007$,+.0039] & .282 \\
\bottomrule
\end{tabular}
\end{table*}

\begin{table}[t]
\centering
\caption{One-vector integration versus $N+1$-request fusion at $N=8$. \ndcg averages T19/T20/BEIR-14; \success averages LoTTE. ``ms/q'' is measured index-side latency per query.}
\label{tab:fusion}
\footnotesize
\setlength{\tabcolsep}{3.5pt}
\begin{tabular}{lrrrr}
\toprule
Method & \ndcg & \success & Requests & ms/q \\
\midrule
AnchorQE & .6591 & .7705 & \best{1} & \best{.19} \\
Weighted CombSUM & .6612 & .7705 & 9 & 4.33 \\
Trust-matched RRF & .6498 & .7693 & 9 & 1.28 \\
Anchored max & \best{.6613} & \best{.7717} & 9 & 4.96 \\
\bottomrule
\end{tabular}
\end{table}

\begin{table*}[!t]
\centering
\caption{Reproduced non-LLM QuDAR comparison. SC-AnchorQE outperforms QuDAR-Simple and QuDAR-Confidence in all 20 QE-strategy--benchmark comparisons.}
\label{tab:qudar}
\scriptsize
\setlength{\tabcolsep}{2.6pt}
\begin{tabular}{llccccc}
\toprule
QE strategy & Method & T19 & T20 & Search & Forum & BEIR-14 \\
\midrule
HyDE & SC-AnchorQE & \best{.7366} & \best{.7340} & \best{.7981} & \best{.7613} & \best{.5516} \\
HyDE & QuDAR-Simple (Equal) & .6332 & .6327 & .7586 & .7199 & .5388 \\
HyDE & QuDAR-Confidence & .6402 & .6341 & .7620 & .7208 & .5393 \\
\addlinespace
Query2Doc & SC-AnchorQE & \best{.7534} & \best{.7397} & \best{.7974} & \best{.7593} & \best{.5487} \\
Query2Doc & QuDAR-Simple (Equal) & .6464 & .6270 & .7520 & .7139 & .5351 \\
Query2Doc & QuDAR-Confidence & .6507 & .6288 & .7538 & .7145 & .5353 \\
\addlinespace
Q2E & SC-AnchorQE & \best{.7339} & \best{.7041} & \best{.7860} & \best{.7546} & \best{.5453} \\
Q2E & QuDAR-Simple (Equal) & .6277 & .5561 & .7325 & .7000 & .5246 \\
Q2E & QuDAR-Confidence & .6297 & .5591 & .7341 & .7002 & .5261 \\
\addlinespace
CoT terms & SC-AnchorQE & \best{.7359} & \best{.7136} & \best{.7896} & \best{.7539} & \best{.5443} \\
CoT terms & QuDAR-Simple (Equal) & .5862 & .5351 & .7230 & .6895 & .5147 \\
CoT terms & QuDAR-Confidence & .5922 & .5382 & .7271 & .6911 & .5165 \\
\bottomrule
\end{tabular}
\end{table*}

\begin{table*}[!t]
\centering
\caption{Cross-generator and cross-retriever analysis. Each cell reports the performance difference (averaged over four QE strategies) $\Delta$ [95\% paired query-bootstrap CI].}
\label{tab:transfer}
\setlength{\tabcolsep}{2pt}
\renewcommand{\arraystretch}{.88}
\resizebox{\textwidth}{!}{%
\begin{tabular}{lccccc}
\toprule
Generator + retriever & T19 & T20 & Search & Forum & BEIR-14 \\
\midrule
\multicolumn{6}{c}{\textit{SC-AnchorQE $-$ DR baseline}} \\
\midrule
Qwen3-8B + BGE & $+.0763\,[+.0268,+.1465]$ & $+.0196\,[+.0017,+.0360]$ & $+.0150\,[+.0099,+.0203]$ & $+.0104\,[+.0071,+.0135]$ & $+.0088\,[+.0067,+.0110]$ \\
Qwen3-8B + GTE & $+.0437\,[+.0019,+.1093]$ & $+.0069\,[-.0039,+.0182]$ & $+.0067\,[+.0025,+.0108]$ & $+.0084\,[+.0055,+.0112]$ & $+.0060\,[+.0042,+.0080]$ \\
Qwen3-8B + Qwen3-Emb & $+.0580\,[+.0159,+.1234]$ & $+.0319\,[+.0143,+.0483]$ & $+.0183\,[+.0130,+.0237]$ & $+.0150\,[+.0117,+.0183]$ & $+.0119\,[+.0103,+.0137]$ \\
Llama-3.1-8B + BGE & $+.0585\,[+.0126,+.1279]$ & $+.0140\,[-.0034,+.0316]$ & $+.0085\,[+.0030,+.0141]$ & $+.0054\,[+.0022,+.0087]$ & $+.0107\,[+.0086,+.0129]$ \\
Llama-3.1-8B + GTE & $+.0497\,[+.0086,+.1150]$ & $+.0035\,[-.0098,+.0176]$ & $+.0039\,[-.0005,+.0082]$ & $+.0037\,[+.0008,+.0066]$ & $+.0059\,[+.0039,+.0079]$ \\
Llama-3.1-8B + Qwen3-Emb & $+.0502\,[+.0059,+.1162]$ & $+.0309\,[+.0153,+.0476]$ & $+.0117\,[+.0063,+.0171]$ & $+.0109\,[+.0076,+.0142]$ & $+.0133\,[+.0116,+.0151]$ \\
Qwen3-1.7B + BGE & $+.0282\,[+.0058,+.0573]$ & $+.0072\,[-.0041,+.0187]$ & $+.0056\,[+.0011,+.0101]$ & $+.0002\,[-.0027,+.0031]$ & $+.0038\,[+.0020,+.0057]$ \\
Qwen3-1.7B + GTE & $+.0143\,[+.0003,+.0294]$ & $-.0069\,[-.0163,+.0021]$ & $+.0002\,[-.0036,+.0039]$ & $+.0008\,[-.0017,+.0033]$ & $+.0014\,[-.0001,+.0029]$ \\
Qwen3-1.7B + Qwen3-Emb & $+.0142\,[+.0007,+.0289]$ & $+.0190\,[+.0070,+.0315]$ & $+.0065\,[+.0022,+.0108]$ & $+.0022\,[-.0004,+.0048]$ & $+.0040\,[+.0026,+.0054]$ \\
\midrule
\multicolumn{6}{c}{\textit{SC-AnchorQE $-$ development-selected fixed AnchorQE}} \\
\midrule
Qwen3-8B + BGE & $+.0194\,[+.0068,+.0358]$ & $+.0018\,[-.0056,+.0091]$ & $+.0041\,[+.0017,+.0065]$ & $+.0022\,[+.0005,+.0038]$ & $+.0017\,[+.0007,+.0026]$ \\
Qwen3-8B + GTE & $+.0424\,[+.0040,+.1065]$ & $+.0024\,[-.0067,+.0122]$ & $+.0061\,[+.0028,+.0095]$ & $+.0054\,[+.0030,+.0078]$ & $+.0036\,[+.0022,+.0050]$ \\
Qwen3-8B + Qwen3-Emb & $+.0051\,[+.0006,+.0104]$ & $+.0025\,[+.0007,+.0045]$ & $+.0020\,[+.0008,+.0034]$ & $+.0007\,[-.0002,+.0015]$ & $+.0015\,[+.0011,+.0019]$ \\
Llama-3.1-8B + BGE & $+.0261\,[+.0060,+.0553]$ & $+.0063\,[-.0085,+.0216]$ & $+.0058\,[+.0012,+.0104]$ & $+.0036\,[+.0008,+.0065]$ & $+.0071\,[+.0053,+.0090]$ \\
Llama-3.1-8B + GTE & $+.0445\,[+.0068,+.1095]$ & $+.0013\,[-.0103,+.0143]$ & $+.0031\,[-.0004,+.0067]$ & $+.0014\,[-.0011,+.0038]$ & $+.0036\,[+.0021,+.0051]$ \\
Llama-3.1-8B + Qwen3-Emb & $+.0132\,[+.0001,+.0308]$ & $+.0020\,[-.0030,+.0072]$ & $+.0017\,[-.0008,+.0042]$ & $+.0016\,[+.0002,+.0030]$ & $+.0033\,[+.0026,+.0040]$ \\
Qwen3-1.7B + BGE & $+.0169\,[+.0004,+.0419]$ & $+.0035\,[-.0030,+.0102]$ & $+.0034\,[+.0004,+.0065]$ & $-.0005\,[-.0024,+.0013]$ & $+.0011\,[-.0001,+.0024]$ \\
Qwen3-1.7B + GTE & $+.0098\,[-.0014,+.0218]$ & $-.0065\,[-.0145,+.0011]$ & $-.0003\,[-.0034,+.0028]$ & $-.0004\,[-.0025,+.0017]$ & $+.0005\,[-.0007,+.0017]$ \\
Qwen3-1.7B + Qwen3-Emb & $+.0032\,[-.0028,+.0105]$ & $+.0048\,[+.0012,+.0087]$ & $+.0022\,[+.0002,+.0043]$ & $+.0001\,[-.0011,+.0013]$ & $+.0005\,[+.0001,+.0010]$ \\
\bottomrule
\end{tabular}}
\end{table*}

The evaluation proceeds in two stages that answer different questions. Section~\ref{sec:integration-diagnostic} selects a fixed interpolation factor on a separate development set so that the benchmark comparison can isolate whether integration alone changes the outcome of QE. Having established that it does, Section~\ref{sec:online-results} removes development labels from factor estimation and evaluates SC-AnchorQE under the online prefix-to-future protocol.

\subsection{Integration Method Matters}
\label{sec:integration-diagnostic}

Table~\ref{tab:main} shows a diagnostic evaluation that compares multiple integration strategies under exact same sets of expansion texts. All rows use the same Qwen3-8B outputs and frozen BGE retriever, only the integration method changes. The conventional published procedures and development-tuned text re-encoding each fall below the DR baseline in 16/20 QE-strategy--benchmark comparisons. In contrast, \method at $\alpha=.15$ exceeds the DR baseline and the conventional procedure in all 20 comparisons; its relative gain over the conventional procedure ranges from .46\% for Query2Doc on T19 to 12.89\% for CoT terms on BEIR-14. Thus changing only integration reverses the dominant result without requiring a better generator.
Appendix~\ref{sec:additional-protocol} expands the BEIR-14 column by collection: AnchorQE improves on published integration in 53/56 strategy--collection comparisons, with at least three of four QE strategies improving on every collection.

\subsection{SC-AnchorQE: Primary Online Results}
\label{sec:online-results}

The above diagnostic experiment identifies the integration problem but used an interpolation factor tuned on the development partitions to choose $\alpha$. SC-AnchorQE addresses the remaining research question by estimating $\alpha$ from unlabeled retrieval signals. Table~\ref{tab:prefix} shows that SC-AnchorQE outperforms the published-integration baseline in all 20 QE-strategy--benchmark comparisons. The largest relative gain over the corresponding conventional procedure is 13.03\% for CoT terms on BEIR-14.

Table~\ref{tab:inference} compares SC-AnchorQE with three baselines: the DR baseline without query expansion, the published-integration baseline that pairs each QE strategy with its original integration method, and fixed-factor AnchorQE with $\alpha=.15$ selected on development partition. We report HyDE, Query2Doc, Q2E, and CoT terms separately. In table~\ref{tab:inference}, "Positive groups" and "query CI" are counts out of five reporting groups. W/L/T counts 60,226 unique queries per strategy, and collection deltas weight 18 collections equally. Brackets are 95\% collection-bootstrap intervals.

Against the DR baseline, SC-AnchorQE improves in 19/20 strategy--group comparisons, with positive query-level CIs in 18/20; all four collection-level CIs also exclude zero. Against published integration, it improves all 20 groups and wins substantially more than it loses for every QE strategy. Gains over fixed AnchorQE are smaller but remain positive in 17/20 group means: the collection-level CI excludes zero for HyDE and Query2Doc, while the Q2E and CoT intervals cross zero. Thus, SC-AnchorQE can successfully estimate interpolation factor without labeled data, and consistently improves retrieval performance across diverse collections.

\subsection{Serving Cost and Linear Fusion}
\label{sec:fusion}

Proposition~1 demonstrates that AnchorQE and weighted CombSUM have the same exact ranking in a shared dense space. We test this with eight saved stochastic HyDE expansions per query, uniform expansion weights, $\alpha=.10$, depth-1,000 retrieval, and the same frozen BGE HNSW or exact inner-product index used by the main evaluation. AnchorQE and weighted CombSUM have .9998 top-10 overlap; their normalized query vectors are identical in the stored floating-point computation (maximum coordinate difference 0). Table~\ref{tab:fusion} shows the practical difference: AnchorQE compiles the linear objective into one index request, whereas separate fusion retrieves the query and eight expansions independently and merges nine candidate lists. Anchored max uses the nonlinear score $s(d)=(1-\alpha)s_q(d)+\alpha\max_i s_{z_i}(d)$; it is slightly stronger but cannot be compiled into one linear query vector. Latency is index-side only and excludes generation and encoding.

\subsection{Comparison with QuDAR}
\label{sec:qudar}

QuDAR \cite{kim-etal-2026-qudar} fuses original and expanded queries across sparse and dense retrievers. We compare its non-LLM Simple and Confidence routes at matched LLM cost, following the original paper and released implementation; Appendix~\ref{sec:additional-protocol} gives reproduction details.

Table~\ref{tab:qudar} shows that SC-AnchorQE is consistently stronger in this fixed-expansion setting, indicating that query-anchored integration transfers more robustly than the compared heterogeneous fusion rules.

\subsection{Stability across Models}

Table~\ref{tab:transfer} evaluates SC-AnchorQE over the complete three-generator by three-retriever matrix. For every generator--retriever configuration, we first average the four QE-strategy scores per query and then report results separately on each benchmark group. The paired bootstrap preserves the collection-level macro-averaging used for LoTTE and BEIR.

Across the 45 generator--retriever--benchmark comparisons, SC-AnchorQE performs better than the DR baseline in 44 cases and outperforms development-selected fixed AnchorQE in 41 cases. The corresponding confidence interval excludes zero on the positive side in 34 and 26 cases, respectively, while no interval is significantly negative. Transfer is strongest for the two 8B generators. Qwen3-1.7B serves as a smaller generator to stress-test the method, acting as a proxy for lower expansion quality. SC-AnchorQE remains above DR baseline in a majority of comparisons, showing that query anchoring often remains robust even when expansions weaken. Its lower coverage than the 8B generators also exposes the limit: integration can control expansion influence, but the quality of generated evidence still sets the attainable gain.

\section{Analysis}
\label{sec:analysis}

\subsection{Why Conjunctive Calibration?}

Equation~\ref{eq:stream-alpha} encodes a direct intuition: expansion trust should be high only when both retrieval strength and agreement with original-query evidence are high. Table~\ref{tab:signals} compares this choice with other label-free estimation methods under the same $B=8$ stream calibration protocol. Four direct agreement rules replace $r_{\mathrm{sup}}$ with clipped query--expansion cosine, top-10 overlap $|D^q_{10}\cap D^z_{10}|/10$, Jaccard, or RBO@10 ($p=.9$), while retaining $r_{\mathrm{top1}}$ as the strength term. Two distributional rules replace agreement with expansion-score concentration: $1-H(p_z)/\log 10$ for entropy or $\mathrm{CV}/(1+\mathrm{CV})$ for score dispersion. The top-score rule instead uses the normalized expansion advantage $[s_{z,1}-s_{q,1}]_+/(s_{z,1}+s_{q,1})$. None is rescaled or selected with benchmark labels.

We also replace the product by the minimum, geometric mean, or arithmetic mean of $r_{\mathrm{top1}}$ and $r_{\mathrm{sup}}$, and evaluate the same product per query rather than per stream. Among stream-level rules, the conjunctive product has the broadest improvement over fixed AnchorQE (17/20 QE-strategy--benchmark comparisons) and the largest median gain. Embedding and ranked-list agreement are useful but less consistent, while the three less-conservative combination functions lose coverage. Per-query factors reach 18/20 positive comparisons but beat the stream product in only 10/20 and require the two probe retrievals for every future query. The stream product therefore offers the strongest deployable balance of effectiveness, conservative trust, and one-request serving after an eight-query prefix.

\begin{table*}[t]
\centering
\caption{Unsupervised factor-rule ablation under the natural $B=8$ prefix-to-future protocol. Each QE-strategy cell counts benchmark groups improved over fixed AnchorQE at $\alpha=.15$; Total is out of 20. Median $\Delta$ is the median native-metric difference over the same 20 separated comparisons. ``2/query'' means that calibration probes continue on every future query.}
\label{tab:signals}
\scriptsize
\setlength{\tabcolsep}{4pt}
\begin{tabular}{lccccccc}
\toprule
Factor rule & HyDE & Query2Doc & Q2E & CoT terms & Total & Median $\Delta$ & Future probes \\
\midrule
Conjunctive product & 5/5 & 5/5 & 4/5 & 3/5 & \best{17/20} & \best{+.0030} & None \\
\midrule
Query--expansion cosine & 5/5 & 4/5 & 3/5 & 1/5 & 13/20 & +.0025 & None \\
Top-10 overlap & 5/5 & 5/5 & 2/5 & 3/5 & 15/20 & +.0020 & None \\
Top-10 Jaccard & 5/5 & 5/5 & 2/5 & 0/5 & 12/20 & +.0004 & None \\
RBO@10 & 5/5 & 5/5 & 3/5 & 2/5 & 15/20 & +.0013 & None \\
Top-score advantage & 0/5 & 0/5 & 0/5 & 0/5 & 0/20 & $-.0093$ & None \\
Entropy concentration & 0/5 & 0/5 & 0/5 & 0/5 & 0/20 & $-.0102$ & None \\
Score dispersion (CV) & 0/5 & 0/5 & 0/5 & 0/5 & 0/20 & $-.0088$ & None \\
\midrule
Minimum & 4/5 & 3/5 & 2/5 & 1/5 & 10/20 & $-.0010$ & None \\
Geometric mean & 3/5 & 3/5 & 2/5 & 1/5 & 9/20 & $-.0011$ & None \\
Arithmetic mean & 3/5 & 3/5 & 2/5 & 1/5 & 9/20 & $-.0011$ & None \\
\midrule
Per-query product & 5/5 & 5/5 & 4/5 & 4/5 & \best{18/20} & +.0027 & 2/query \\
\bottomrule
\end{tabular}
\end{table*}

\subsection{Qualitative Analysis: How the Query Anchor Changes Retrieval}

Table~\ref{tab:cases} shows four concrete examples of what the query anchor can and cannot do. In the first two, anchoring blocks a wrong answer or adds useful details without replacing the query. In the last two, limited movement cannot repair irrelevant generated text, and one stream-level factor may give too little weight to an unusually useful expansion.

\begin{table*}[t]
\centering
\caption{Illustrative cases from the fixed-expansion BGE comparison. Scores are per-query \ndcg for BEIR and \success for LoTTE. ``Conventional'' is the corresponding expansion-only or text re-encoding method.}
\label{tab:cases}
\footnotesize
\setlength{\tabcolsep}{3pt}
\begin{tabular}{p{.15\textwidth}p{.31\textwidth}ccc p{.22\textwidth}}
\toprule
Query / source & Saved expansion (verbatim) & DR baseline & Conventional & AnchorQE & Interpretation \\
\midrule
Actor in \emph{Arrival} nominated for \emph{The Town} / HotpotQA &
``John David Washington'' &
\best{.9197} & .0000 & \best{.9197} &
Expansion-only follows a wrong entity; anchoring preserves the original
evidence path. \\
\addlinespace
Best chopping-board material for meat / LoTTE Lifestyle &
``Wood, plastic, stainless steel, bamboo, silicone, non-slip, food-safe,
durability, hygiene, bacteria resistance, ease of cleaning, cutting
performance, heat resistance, knife maintenance, environmental impact,
cost-effectiveness'' &
.0 & .0 & \best{1.0} &
A bounded correction adds useful facets without adopting the full re-encoded
direction. \\
\addlinespace
Most surprising result in mathematics / LoTTE Science &
``G\"odel's incompleteness theorems, Riemann Hypothesis, Mandelbrot set, Four
color theorem, Cantor's diagonal argument, Fermat's Last Theorem, Banach--Tarski
paradox, Poincar\'e conjecture, Euler's identity, Hilbert's problems'' &
\best{1.0} & .0 & .0 &
Limited movement cannot make irrelevant evidence useful. \\
\addlinespace
Standard meridian of India / Natural Questions &
``Standard Meridian of India 82 30 E 82.5 East Longitude Indian Standard Time
Meridian 82 30 East Longitude 82.5 E Meridian India Standard Time Meridian India
Time Meridian 82 30 E Meridian India 82.5'' &
.0 & \best{1.0} & .3333 &
A stream-level factor can underweight an unusually strong expansion. \\
\bottomrule
\end{tabular}
\end{table*}

\section{Conclusion}

Our main finding is simple: a generated expansion is only useful if it is integrated well. The same expansion can help or hurt retrieval depending on how it is used. Thus, integration is a critical part of the QE method rather than a minor implementation choice. \method provides a direct solution by keeping the original query and the expansion separate until the vector stage, where an explicit interpolation factor controls expansion trust. SC-AnchorQE estimates this factor without supervision while preserving single-vector retrieval over the original, unchanged index. This framework is training-free, works across generators and retrievers, and limits how easily generated text can pull retrieval away from the original query. 
Our results show that \method is both \emph{effective}---outperforming conventional integration methods---and \emph{robust}---outperforming pure dense retrieval without QE in nearly all evaluated settings. These benefits generalize across generators and retrievers, expansion strategies, and datasets.

\section{Limitations}

SC-AnchorQE assumes that a short unlabeled prefix represents a reasonably stable future stream; rapid distribution shifts can make its stream-level factor stale, and one shared factor cannot handle every unusually strong or misleading expansion. Transfer also depends on generation quality, so better integration cannot recover information absent from weak expansions. Our benchmark protocol is a strict no-lookahead simulation rather than temporal traffic, and broader deployment studies remain necessary. Finally, the fusion equivalence assumes a shared dense space with raw dot products and exact ranking; prefix calibration adds probe requests, although subsequent retrieval uses one index request and generation remains the dominant cost.

\newpage
\appendix
\section{Additional Protocol Details}
\label{sec:additional-protocol}

\begin{table*}[tp]
\centering
\caption{Per-collection expansion of the BEIR-14 diagnostic in Table~\ref{tab:main}. The DR baseline is shared across QE strategies. Each remaining cell gives fixed-factor AnchorQE \ndcg followed in parentheses by its absolute delta versus the corresponding published integration method.}
\label{tab:beir-datasets}
\scriptsize
\setlength{\tabcolsep}{4pt}
\begin{tabular}{lccccc}
\toprule
Dataset & DR baseline & HyDE & Query2Doc & Q2E & CoT terms \\
\midrule
TREC-COVID & .7436 & .7692 $(+.0481)$ & .7676 $(+.0002)$ & .7630 $(-.0071)$ & .7645 $(+.0215)$ \\
NFCorpus & .3819 & .3913 $(+.0065)$ & .3912 $(+.0117)$ & .3896 $(+.0065)$ & .3875 $(+.0272)$ \\
FiQA & .4492 & .4641 $(+.0547)$ & .4603 $(+.0437)$ & .4514 $(+.0469)$ & .4570 $(+.0496)$ \\
ArguAna & .4580 & .4604 $(+.0457)$ & .4607 $(+.0052)$ & .4606 $(+.0118)$ & .4626 $(+.0133)$ \\
Touch\'e 2020 & .2539 & .2525 $(+.0340)$ & .2498 $(-.0079)$ & .2480 $(+.0074)$ & .2466 $(+.0191)$ \\
SCIDOCS & .2261 & .2289 $(+.0163)$ & .2291 $(+.0028)$ & .2300 $(+.0129)$ & .2285 $(+.0122)$ \\
Quora & .8898 & .8885 $(+.1830)$ & .8891 $(+.0690)$ & .8885 $(+.0379)$ & .8890 $(+.0505)$ \\
DBPedia-Entity & .4410 & .4551 $(+.0479)$ & .4500 $(+.0564)$ & .4492 $(+.0898)$ & .4459 $(+.1208)$ \\
Climate-FEVER & .3657 & .3704 $(+.0989)$ & .3695 $(+.0970)$ & .3670 $(+.1071)$ & .3675 $(+.1124)$ \\
NQ & .5502 & .5821 $(+.0474)$ & .5771 $(+.0461)$ & .5657 $(+.0656)$ & .5659 $(+.0954)$ \\
HotpotQA & .7416 & .7522 $(+.1157)$ & .7517 $(+.0765)$ & .7460 $(+.1331)$ & .7447 $(+.1366)$ \\
SciFact & .7463 & .7510 $(-.0053)$ & .7533 $(+.0030)$ & .7537 $(+.0084)$ & .7535 $(+.0073)$ \\
FEVER & .8719 & .8826 $(+.0840)$ & .8803 $(+.0869)$ & .8732 $(+.1817)$ & .8728 $(+.1754)$ \\
CQADupStack & .4223 & .4318 $(+.0545)$ & .4305 $(+.0201)$ & .4265 $(+.0227)$ & .4252 $(+.0277)$ \\
\bottomrule
\end{tabular}
\end{table*}

\paragraph{BEIR-14 breakdown.} Table~\ref{tab:beir-datasets} expands the fixed-expansion diagnostic in Table~\ref{tab:main}. AnchorQE improves over published integration in 53/56 QE-strategy--collection comparisons and over development-tuned text re-encoding in all 56. Every collection improves for at least three of four QE strategies, so the BEIR-14 macro reflects a broad pattern rather than gains concentrated in a few collections. CQADupStack is first macro-averaged over its 12 domains, as in the main evaluation.

\paragraph{Expansion encoding role.} AnchorQE treats a generated expansion as query-side retrieval evidence and therefore encodes it with $E_q$ in Eq.~\ref{eq:anchorqe}. Table~\ref{tab:eq-ed} tests the alternative of encoding the expansion with $E_d$ while holding the saved max128 expansion, $E_q$-encoded query anchor, frozen BGE index, and $\alpha=.15$ fixed. We report HyDE and Query2Doc separately because both generate document-like text and thus make the $E_q$ versus $E_d$ choice most salient. $E_q$ is better in 6/10 QE-strategy--benchmark comparisons, but every absolute difference is at most .0038. The method is therefore insensitive to this encoding-role choice in these conditions; we retain $E_q$ because the expansion ultimately modifies a query vector.

\begin{table}[tbp]
\centering
\caption{Expansion-encoder ablation with max128 Qwen3-8B generations, BGE, and fixed $\alpha=.15$. Only the saved expansion's encoding role changes. HyDE and Query2Doc are reported separately; bold marks the better encoder within each QE strategy and benchmark.}
\label{tab:eq-ed}
\small
\setlength{\tabcolsep}{1.5pt}
\begin{tabularx}{\columnwidth}{@{}l>{\raggedright\arraybackslash}Xccccc@{}}
\toprule
QE strategy & Expansion encoder & T19 & T20 & Search & Forum & BEIR-14 \\
\midrule
HyDE & $E_q$ & \best{.7154} & .7303 & \best{.7913} & .7573 & \best{.5486} \\
HyDE & $E_d$ & .7147 & \best{.7320} & .7909 & \best{.7576} & .5481 \\
\addlinespace
Query2Doc & $E_q$ & \best{.7300} & .7299 & \best{.7924} & \best{.7569} & .5472 \\
Query2Doc & $E_d$ & .7263 & \best{.7306} & .7923 & .7555 & \best{.5473} \\
\bottomrule
\end{tabularx}
\end{table}

\paragraph{Alternative generation-length regimes.} Table~\ref{tab:length-regimes} records the main fixed-expansion, full-query diagnostic under the original 64-token cap and open-ended EOS termination. Each regime uses its own development-selected fixed interpolation factor.

\begin{table}[tbp]
\centering
\caption{Major results for the max64 and open-ended generation regimes.}
\label{tab:length-regimes}
\small
\setlength{\tabcolsep}{1.25pt}
\begin{tabularx}{\columnwidth}{@{}>{\raggedright\arraybackslash}p{.16\columnwidth}>{\raggedright\arraybackslash}Xccccc@{}}
\toprule
Regime & Integration method & T19 & T20 & Search & Forum & BEIR-14 \\
\midrule
max64 & DR baseline & .6765 & .7056 & .7780 & .7470 & .5387 \\
& Conventional published & .6943 & .6716 & .7429 & .7012 & .4904 \\
& Tuned text re-encoding & .7124 & .6745 & .7632 & .7193 & .5135 \\
& AnchorQE, fixed $\alpha=.10$ & \best{.7144} & \best{.7133} & \best{.7867} & \best{.7522} & \best{.5436} \\
\addlinespace
Open-ended & DR baseline & .6765 & .7056 & .7780 & .7470 & .5387 \\
& Conventional published & .6862 & .6699 & .7461 & .7040 & .4925 \\
& Tuned text re-encoding & .7108 & .6725 & .7633 & .7198 & .5128 \\
& AnchorQE, fixed $\alpha=.15$ & \best{.7229} & \best{.7214} & \best{.7885} & \best{.7552} & \best{.5456} \\
\bottomrule
\end{tabularx}
\end{table}

\paragraph{Prefix-size sensitivity.}
SC-AnchorQE is stable across the tested prefix sizes. As shown in Table~\ref{tab:prefix-sensitivity}, the largest reporting-group spread across $B=8,16,32$ is only .0027.

\begin{table}[!h]
\centering
\caption{Full-query four-strategy macros across calibration-prefix sizes.}
\label{tab:prefix-sensitivity}
\footnotesize
\renewcommand{\arraystretch}{.9}
\begin{tabular}{lccccc}
\toprule
$B$ & T19 & T20 & Search & Forum & BEIR-14 \\
\midrule
8 & .7399 & .7228 & .7928 & .7573 & .5475 \\
16 & .7375 & .7229 & .7930 & .7574 & .5475 \\
32 & .7373 & .7228 & .7929 & .7574 & .5475 \\
\bottomrule
\end{tabular}
\end{table}

\paragraph{QuDAR reproduction details.} We use the released QuDAR fusion implementation at commit \texttt{0702721e}. The dense streams use the same frozen BGE indexes as AnchorQE. The sparse matrix backend is algebraically equivalent to QuDAR's BM25Okapi configuration (lowercase whitespace tokenization, English stemming, $k_1=1.5$, $b=.75$, $\epsilon=.25$) while avoiding an infeasible Python-object representation for 8.8M MS MARCO passages. Every QE strategy is evaluated independently.

\section{Exact Generation Prompts}

The templates below implement the expansion objects introduced by HyDE, Query2Doc, and prompted QE \cite{gao2023hyde,wang2023query2doc,jagerman2023query,zhang-etal-2024-exploring-best}. Every strategy uses the user message \texttt{Query: \{query\}}.

\paragraph{HyDE \cite{gao2023hyde}.} ``For a given search query, write a brief, factual, and professional hypothetical document that directly answers or contains the relevant information. Output only the passage.''

\paragraph{Query2Doc \cite{wang2023query2doc}.} ``Write one concise pseudo-document that a relevant corpus passage could contain. Include important entities, synonyms, and answer facets, but do not invent narrow facts not implied by the query. Output only the pseudo-document.''

\paragraph{Q2E \cite{jagerman2023query}.} ``Convert the query into a compact expansion query with related terms, synonyms, aliases, acronyms, and key entities. Output only space-separated terms or short phrases.''

\paragraph{CoT terms \cite{jagerman2023query,zhang-etal-2024-exploring-best}.} ``Internally decompose the query into intent, entities, constraints, and likely answer facets, then output only the final expansion terms. Do not show the reasoning.''

\section{Development Tuning of the Text Re-encoding Baseline}
\label{sec:text-dev-tuning}

The development-tuned text baseline searches over different ways of composing the original query and its saved expansion before passing the combined text through the frozen query encoder. This development tuning changes only the text-composition recipe.

We construct the candidate grid from four expansion-length settings, three separators, two text orders, and three query-repetition counts:
\[
4\ \text{lengths}\times3\ \text{separators}\times2\ \text{orders}\times3\ \text{repetition counts}=72.
\]
The expansion-length settings retain expansion tokens up to (i) the token length of the original query, (ii) 32 tokens, (iii) 64 tokens, or (iv) the complete saved expansion, subject to the encoder's 512-token limit. The separators are a space, \texttt{[SEP]}, and a newline. The two orders place either the query or the expansion first. The original query is repeated 1, 2, or 5 times before composition.

Each QE strategy selects its recipe independently on the 6,980-query MS MARCO passage development set. The primary selection metric is \ndcg. No TREC-DL, LoTTE, or BEIR relevance judgments are used for selection.

The selected recipes consistently retain a short expansion and avoid repeating the query. The preferred order differs by expansion type: document-like HyDE and Query2Doc expansions follow the query, whereas term-like Q2E and CoT expansions precede it. This tuning provides a stronger text re-encoding baseline than directly adopting a single published concatenation format, while keeping all test-collection results strictly held out.

\clearpage

\bibliographystyle{ACM-Reference-Format}
\bibliography{references}

\end{document}